# Performance Analysis of Markov Modulated 1-Persistent CSMA/CA Protocols with Exponential Backoff Scheduling

Pui King Wong[1], Dongjie Yin[1], and Tony T. Lee[1,2]

**Abstract.** This paper proposes a Markovian model of 1-persistent CSMA/CA protocols with *K*-Exponential Backoff scheduling algorithms. The input buffer of each access node is modeled as a Geo/G/1 queue, and the service time distribution of each individual head-of-line packet is derived from the Markov chain of the underlying scheduling algorithm. From the queuing model, we derive the characteristic equation of network throughput and obtain the stable throughput and bounded delay regions with respect to the retransmission factor. Our results show that the stable throughput region of the exponential backoff scheme exists even for an infinite population. Moreover, we find that the bounded delay region of exponential backoff is only a sub-set of its stable throughput region due to the large variance of the service time of input packets caused by the capture effect. All analytical results presented in this paper are verified by simulations.

## 1 Introduction

Carrier Sense Multiple Access (CSMA) is a standardized Media Access Control (MAC) protocol for the transmission of network stations over a commonly shared medium, such as 802.11 wireless LANs or Ethernet networks. It is a set of rules for resolving collisions when two or more nodes attempt to use a transmission channel simultaneously. The network stations governed by CSMA with collision avoidance (CSMA/CA) check the busy/idle status of the transmission channel before attempting to transmit. If the channel is detected busy, the packets involved in the collision attempt to retransmit after a random time interval. With exponential backoff scheduling, this random waiting time interval for retransmission geometrically increases with the number of collisions encountered by the packet.

The *p*-persistent CSMA protocols try to maximize the channel utilization by continuously monitoring the transmission medium. An active node with a packet ready to transmit constantly checks the channel status. If the channel is sensed idle, the node transmits the packet with a probability *p*. Otherwise, the node will persistently monitor the channel until it becomes idle, and then transmits again with the same probability *p*. This process repeats until the packet is dispatched. Collisions may still occur when two or more packets are simultaneously transmitted. The 1-Persistent Carrier-Sense Multiple Access (1P-CSMA) protocol is a special case of the *p*-persistent CSMA protocol with *p* = 1.

1. Dept. of Information Engineering, The Chinese University of Hong Kong, Hong Kong (email: {wpk007, ydj008, ttlee}@ie.cuhk.edu.hk)
2. Dept. of Electronics Engineering, Shanghai Jiao Tong University, Shanghai, China.

The complete analysis of a buffered random-access network has long been considered to be a challenging problem even for the simplest case. The throughput analysis of slotted 1-persistent CSMA for a network with an infinite population of stations was first investigated in [1] by Kleinrock and Tobagi. Many follow-up researches were proposed to analyze the performance of 1-persistent CSMA protocol. For example, a finite population system with the slotted 1-persistent CSMA, which can also be extended to infinite population cases, was analyzed in [2]. The characteristic equation for network throughput was re-derived in [3] by using a state-transition diagram of the transmission channel. However, these results are limited to the maximum throughput analysis of a saturated network in which the scheduling of collided packets was not considered.

The *K*-Exponential Backoff algorithm is a widely used collision resolution scheme in most MAC protocols such as the 802.11 [4]-[8]. A packet encountered collisions $i$ times is scheduled for retransmission with probability $q^i$, for $i = 1,…, K$. The parameter $q$ is called the retransmission factor with $0 < q < 1$ and $K$ is the cut-off phase. Many published works [9]-[13] studied the performance of MAC protocols in conjunction with the exponential backoff algorithm under a saturation traffic condition. A two-dimensional Markov chain that described the exponential backoff algorithm was proposed by Bianchi in [9] to obtain the network throughput of the 802.11 Distributed Coordination Function (DCF) protocol. Authors in [10]-[13] applied the two-dimensional Markov chain to analyze the performance of network throughput of the 802.11 DCF under different scenarios. Although these works, together with the previous results in [1]-[3], have revealed the connection between the maximum throughput and the underlying backoff algorithm, issues on throughput stability and bounded delay remain open under the saturation assumption. Some non-saturated analyses of 802.11 protocols under a simplified buffer-less assumption were reported in [14]-[16], in which the queuing behaviors of input buffers were completely ignored.

The aim of this paper is to provide a complete performance analysis of the 1P-CSMA protocol under a non-unsaturated traffic environment. Our approach is an extension of the Markov model for non-persistent CSMA networks proposed in [17], which includes both the stable throughput region and the corresponding delay analysis of NP-CSMA. The throughput of a buffered random-access network is determined by the activities of Head-Of-Line (HOL) packets in the entire network. In this paper, we assume that the input to each node is a Bernoulli process with rate $\lambda$ packet/timeslot and each input buffer is modeled as a Geo/G/1 queue. The service time of HOL packets is described by a Markov chain that represents the backoff scheduling algorithm from which the steady-state network throughput can be expressed as a function of the aggregate attempt rate $G$ of HOL packets. Using an extension of the methodology developed in [17], the complete queuing analysis enables us to specify the stable regions of both throughput and bounded mean delay for 1P-CSMA. Specifically, we characterize the stable throughput region of retransmission factor $q$ from the first moment of packet service time with a given aggregate input rate $\hat{\lambda} = n\lambda$ packet/timeslot, where $n$ is the number of nodes in the system. Furthermore, the Pollaczek-Khinchin (P-K) formula of Geo/G/1 queue reveals that a bounded mean delay requires a bounded second moment of service time, which also depends on the retransmission scheduling algorithm employed in 1P-CSMA protocols.

The exponential backoff scheduling algorithm with infinite cutoff phase ($K = \infty$) is of special interest in practical operations of 1P-CSMA protocols. Our result on the throughput of exponential backoff scheme agrees with that reported in [4] that the network can have non-zero throughput even if the number of nodes $n \to \infty$. Moreover, the bounded delay region determined by our queuing model is only a subset of the stable throughput region. For any retransmission factor $q$ inside the stable throughput region but not within the bounded delay region, we show that a stable throughput of the entire network can be achieved with a large delay variance due to the "capture effect" described in [18] and [19]. Thus, the maximum throughput of the network within the bounded delay region is slightly smaller than that achievable in the stable throughput region.

The rest of this paper is organized as follows. The channel model and input buffer model that characterize the network throughput of 1P-CSMA protocols are presented in section 2 and section 3, respectively. Section 4 describes the general requirements of stable conditions. Based on these conditions, the stable throughput and bounded delay regions of the 1P-CSMA protocol with exponential backoff scheme are specified in section 5. Section 6 provides a conclusion.

## 2 Markov chain of Slotted 1P-CSMA

In a slotted 1P-CSMA network, the time axis is slotted and the network is synchronized. Packets can be sent only at the beginning of a timeslot, and each packet is identical and takes one slot time for transmission. Suppose the ratio of propagation delay to packet transmission time is $a$, then the timeslot can be further divided into mini-slots of slot size $a$. According to the channel status displayed in Fig. 1, the busy period is a series of transmission periods each with $1 + a$ timeslots, and pure idle period is the time in which the channel is idle and no packet presents awaiting transmission. Then the time axis can be considered as a sequence of alternating busy periods and pure idle periods.

There are two types of transmissions with different probabilities of success. The leading transmission period in each busy period is called the Type I transmission period, in which the packet is sent successfully if, and only if, no one is scheduled to access the channel in the previous mini-slot. All subsequent transmission periods are defined as Type II transmissions. Due to the persistence property of 1P-CSMA, packets scheduled to access the channel in any transmission period accumulate at the end of the previous transmission period; hence, the second type of transmission is successful if, and only if, no one attempts to access the channel during the previous transmission period.

In each transmission period, a packet is either successfully transmitted or collided. Therefore, the channel has three fundamental states: pure idle (Idle), successful transmission (Suc), and collision (Col). For the 1P-CSMA, the successful transmission state should be divided into two sub-states Suc1 and Suc2 corresponding to the Type I and Type II transmission periods, respectively. The state transitions of the channel can be described by the Markov chain shown in Fig. 2.

The limiting probabilities $\pi_{Idle}$, $\pi_{Suc}$ and $\pi_{Col}$ of the Markov chain satisfy the following set of equations:

$$\begin{aligned}
(1-P_{Idle,Idle}) \cdot \pi_{Idle} &= P_{Suc1,Idle} \cdot \pi_{Suc1} + P_{Suc2,Idle} \cdot \pi_{Suc2} + P_{Col,Idle} \cdot \pi_{Col} \\
\pi_{Suc1} &= P_{Idle,Suc1} \cdot \pi_{Idle} \\
(1-P_{Suc2,Suc2}) \cdot \pi_{Suc2} &= P_{Suc1,Suc2} \cdot \pi_{Suc1} + P_{Col,Suc2} \cdot \pi_{Col} \\
(1-P_{Col,Col}) \cdot \pi_{Col} &= P_{Idle,Col} \cdot \pi_{Idle} + P_{Suc1,Col} \cdot \pi_{Suc1} + P_{Suc2,Col} \cdot \pi_{Suc2}
\end{aligned} \quad (1)$$

Since packets attempt to access the channel only when the channel is detected as idle, the attempt rate in any busy period is zero. During a pure idle period, the aggregate attempts generated by all fresh and re-scheduled HOL packets form a Poisson stream with rate $G$; therefore, the probability that no attempt is generated in a mini-slot is $e^{-aG}$. From a system point of view, the Type I transmission is successful only if there is exactly one attempting packet in a mini-slot with the probability $aGe^{-aG}$.

All packets scheduled to access the channel in any transmission period are accumulated at the end of the previous period, and then attempt to access the channel at the same time. Thus, the transition probability from transmission period to idle period is equivalent to that when there is no scheduled packets within $(1+a)$ slot times, i.e., $e^{-(1+a)G}$. Similarly, the probability of having a successful Type II transmission is $(1+a)Ge^{-(1+a)G}$. Hence, the transition probabilities of the Markov chain are given as follows:

$$\begin{aligned}
P_{Idle,Idle} &= e^{-aG} \\
P_{Idle,Suc1} &= aGe^{-aG} \\
P_{Idle,Col} &= 1 - e^{-aG} - aGe^{-aG} \\
P_{Suc1,Idle} &= P_{Suc2,Idle} = P_{Col,Idle} = e^{-(1+a)G} \\
P_{Suc1,Suc2} &= P_{Suc2,Suc2} = P_{Col,Suc2} = (1+a)Ge^{-(1+a)G} \\
P_{Suc1,Col} &= P_{Suc2,Col} = P_{Col,Col} = 1 - e^{-(1+a)G} - (1+a)Ge^{-(1+a)G}
\end{aligned} \quad (2)$$

We obtain the following limiting probabilities from (1) and (2) in a straightforward manner:

$$\pi_{Idle} = \frac{e^{-(1+a)G}}{1-e^{-aG}+e^{-(1+a)G}}$$
$$\pi_{Suc1} = \frac{aGe^{-aG}e^{-(1+a)G}}{1-e^{-aG}+e^{-(1+a)G}}$$
$$\pi_{Suc2} = \frac{(1+a)Ge^{-(1+a)G}\left(1-e^{-aG}\right)}{1-e^{-aG}+e^{-(1+a)G}}$$
$$\pi_{Col} = \frac{1-e^{-aG}-(1+a)Ge^{-(1+a)G}+Ge^{-aG}e^{-(1+a)G}}{1-e^{-aG}+e^{-(1+a)G}}$$
(3)

The time-average probabilities of each state can be easily obtained from sojourn times of the Idle, Suc, and Col states $t_{Idle}=a$, and $t_{Suc1}=t_{Suc2}=t_{Col}=1+a$, respectively. Of particular importance, the probability that the channel is either in Suc1 or Suc2 states is given by

$$\tilde{\pi}_{Suc1}+\tilde{\pi}_{Suc2} = \frac{(1+a)(\pi_{Suc1}+\pi_{Suc2})}{(1+a)(\pi_{Suc1}+\pi_{Suc2}+\pi_{Col})+a\cdot\pi_{Idle}} = \frac{(1+a)Ge^{-(1+a)G}\left(1+a-e^{-aG}\right)}{(1+a)\left(1-e^{-aG}\right)+ae^{-(1+a)G}}.$$
(4)

Since the successful transmission of a packet only takes $1/(1+a)$ of the time in a transmission period, the following network throughput is defined by the fraction of the time that the channel is productive:

$$\hat{\lambda} = \frac{\tilde{\pi}_{Suc1}+\tilde{\pi}_{Suc2}}{(1+a)} = \frac{Ge^{-(1+a)G}\left(1+a-e^{-aG}\right)}{(1+a)\left(1-e^{-aG}\right)+ae^{-(1+a)G}},$$
(5)

which is consistent with Kleinrock and Tobagi's result in [1]. Although the network throughput can be obtained from the above model, but it is by no means a comprehensive performance analysis of the system because the re-scheduling of collided HOL packets is not considered in the channel model. The next section describes a more detailed queuing model of the input buffer in which the service time distribution is derived in the context of the underlying backoff scheduling algorithm.

## 3  Queuing Model of Input Buffer

The input buffer of each node is modeled as a Geo/G/1 with a Bernoulli arrival process of rate $\lambda$ packets/timeslot. As we mentioned before, we consider the 1P-CSMA protocol with the *K*-Exponential Backoff algorithm for contention resolutions. The behavior of each HOL packet with this backoff algorithm is described by a simple flow chart shown in Fig. 3. A fresh HOL packet is initially in phase 0 and sent only when an idle channel is detected. If the channel is busy, it waits until the channel becomes idle and then transmits the packet immediately. If the transmission is successful, a new fresh HOL packet starts from phase 0. Otherwise, the packet is in backoff mode in which the random waiting time is determined by the current backoff phase. A packet is in phase *i*, if it has encountered collision *i* times. The *K*-Exponential Backoff algorithm allows an HOL packet in phase *i* to retransmit with probability $q^i$, for $i=1,\ldots,K$, where $0<q<1$ is the retransmission factor and *K* is the cut-off phase. That is, the retransmission probability decreases exponentially with the number of collisions, up to *K* times, experienced by the backlogged HOL packets.

Based on the flow chart of the 1P-CSMA protocol shown in Fig. 3, the corresponding state transition diagram with the *K*-Exponential Backoff algorithm is shown in Fig. 4. An HOL packet is in one of the three fundamental states: sensing ($S_i$), transmission, and waiting ($W_i$), $i=0,...,K$, at any time. Note the transmission state is split into two sub-states: $F_i$ and $F_i'$, corresponding to the two types of transmissions, respectively.

If channel idle is detected in the sensing state $S_i$, an HOL packet is sent and moved to state $F_i$ with probability $q^i$. Otherwise, the corresponding node continually monitors the channel activity until an idle channel is sensed. This persistent sensing action is represented by the waiting state $W_i$. When the channel turns idle, the packet moves to the Type II transmission state $F_i'$ with probability $q^i$. If the packet is successfully sent, a fresh HOL packet starts with the initial sensing state $S_0$; otherwise, the collided packet moves to phase $i+1$ and repeats the above process starting from the sensing state $S_{i+1}$. Moreover, for the sake of simplicity, we assume that a new packet at an empty buffer arrives at the beginning of the transmission period if the channel is busy, so that the waiting time is 1 slot time for any phase *i*. The transition probabilities of the Markov chain are derived below.

Let *α* be the probability that the channel is in pure idle state. Then it can be calculated from the limiting probabilities given by (3) as follows:

$$\alpha = \frac{a \cdot \pi_{Idle}}{a \cdot \pi_{Idle} + (1+a)(\pi_{Suc1} + \pi_{Suc2} + \pi_{Col})} = \frac{ae^{-(1+a)G}}{(1+a)(1-e^{-aG}) + ae^{-(1+a)G}}. \quad (6)$$

It is illustrated in Fig. 4 that the Type I transmission is successful only if all other nodes do not attempt to send in the first mini-slot of the busy period. Thus, the probability of a successful Type I transmission is

$$p_1 = e^{-aG}. \quad (7)$$

While the success of the Type II transmission requires that no other nodes attempt to send in the previous transmission period; hence, the probability of a successful Type II transmission is

$$p_2 = e^{-(1+a)G}. \quad (8)$$

Type I transmissions lead busy periods; therefore, the probability that a transmission period is Type I is equivalent to the probability *α* given by (6) that the system is in a pure idle period. It follows that the probability of successful transmission of 1P-CSMA is given by

$$p = \Pr\{\text{Type 1 transmission}\}p_1 + \Pr\{\text{Type 2 transmission}\}p_2$$
$$= \alpha p_1 + (1-\alpha)p_2 \qquad (9)$$
$$= \frac{(1+a-p_1)p_2}{(1+a)(1-p_1)+ap_2}.$$

Substitute (7) and (8) into (9), and the probability of successful transmission can be expressed by the following function of the attempt rate $G$:

$$p(G) = \frac{e^{-(1+a)G}\left(1+a-e^{-aG}\right)}{(1+a)\left(1-e^{-aG}\right)+ae^{-(1+a)G}}. \qquad (10)$$

It should be noted that the probability of success given by (10) is consistent with (5) because $p = \hat{\lambda}/G$ by definition.

Let $s_i$, $f_i$, $f_i'$, and $w_i$ be the respective limiting probabilities of states $S_i$, $F_i$, $F_i'$, and $W_i$ of the Markov chain shown in Fig. 4, from which we obtain the following set of state equations:

$$s_i = \begin{cases} p_1(f_0 + f_1 + \cdots + f_K) + p_2(f_0' + f_1' + \cdots + f_K'), \text{ for } i=0 \\ \alpha(1-q^i)s_i + (1-q^i)w_i + (1-p_1)f_{i-1} + (1-p_2)f_{i-1}', \text{ for } i=1,\ldots,K-1 \\ \alpha(1-q^K)s_K + (1-q^K)w_K + (1-p_1)(f_{K-1}+f_K) + (1-p_2)(f_{K-1}'+f_K'), \text{ for } i=K. \end{cases}$$
$$w_i = (1-\alpha)s_i, \text{ for } i=0,\ldots,K \qquad (11)$$
$$f_i = \alpha q^i s_i, \text{ for } i=0,\ldots,K$$
$$f_i' = q^i w_i, \text{ for } i=0,\ldots,K$$

It can be proven from (11) that if $p + q > 1$, then all states of the Markov chain are positive recurrent and aperiodic. Thus, the time-average probabilities of those states can be determined from (11) with the sojourn time $t_{S_i} = a$, $t_{F_i} = 1$, $t_{F_i'} = 1$, and $t_{W_i} = 1$ of state $S_i$, $F_i$, $F_i'$, and $W_i$, respectively, for $i = 0, \ldots, K$ as follows:

$$\tilde{s}_i = \begin{cases} \frac{a}{D}\left(\frac{1-p}{q}\right)^i, & i=0,1,\ldots,K-1 \\ \frac{a}{pD}\left(\frac{1-p}{q}\right)^K, & i=K \end{cases}$$
$$\tilde{w}_i = \begin{cases} \frac{(1-\alpha)}{D}\left(\frac{1-p}{q}\right)^i, & i=0,1,\ldots,K-1 \\ \frac{(1-\alpha)}{pD}\left(\frac{1-p}{q}\right)^K, & i=K \end{cases}$$
$$\tilde{f}_i = \begin{cases} \frac{\alpha(1-p)^i}{D}, & i=0,1,\ldots,K-1 \\ \frac{\alpha(1-p)^K}{pD}, & i=K \end{cases} \qquad (12)$$
$$\tilde{f}_i' = \begin{cases} \frac{(1-\alpha)(1-p)^i}{D}, & i=0,1,\ldots,K-1 \\ \frac{(1-\alpha)(1-p)^K}{pD}, & i=K \end{cases}$$

where $D = \dfrac{q(1+a-\alpha)}{p+q-1}\left[1-\dfrac{(1-q)}{p}\left(\dfrac{1-p}{q}\right)^{K+1}\right]+\dfrac{1}{p}$.

The offered load $\rho$ of each input queue is the probability that the queue is non-empty. It is the basic measure for analyzing the performance of each input buffer. The input rate $\lambda$ of the Bernoulli arrival process can be interpreted as the probability of finding a packet arrived at input in any time slot. Each input packet will eventually become a fresh HOL packet, and visit the transmission states in phase 0 for one slot time. Therefore, the input rate $\lambda$ should be equal to $\rho(\tilde{f}_0 + \tilde{f}_0')$, the probability of finding a phase 0 HOL packet in either Type I or Type II transmission states in any time slot. With the time-average probability of state $F_0$ and $F_0'$ given in (12), the expression of the offered load can be obtained as follows:

$$\rho = \dfrac{\lambda}{\tilde{f}_0 + \tilde{f}_0'} = \lambda q \dfrac{1+a-\alpha}{p+q-1}\left[1-\dfrac{(1-q)}{p}\left(\dfrac{1-p}{q}\right)^{K+1}\right]+\dfrac{\lambda}{p}. \tag{13}$$

We show in the next theorem that the network throughput derived from the service time of HOL packets is the same as (5), which was previously obtained from the channel model shown in Fig. 2. Although the service time is obviously dependent on the retransmission factor $q$ of the backoff scheduling algorithm, the fact that the throughput is invariant with respect to $q$ implies that the stability of the system cannot be determined by the characteristic equation of throughput alone; it is mainly related to the queuing behavior of each input buffer.

**Theorem 1.** For buffered 1-persistent CSMA with $K$-Exponential Backoff, the throughput in equilibrium is given by

$$\hat{\lambda}_{out} = \hat{\lambda} = \dfrac{-\ln p_1}{a} \cdot p = \dfrac{Ge^{-(1+a)G}(1+a-e^{-aG})}{(1+a)(1-e^{-aG})+ae^{-(1+a)G}}. \tag{14}$$

Proof: A particular HOL packet is ready to be sent only if an idle channel is detected. The probability of successful Type I transmission $p_1$ for a desired node is the conditional probability that none of the other nodes accesses the channel given that all nodes sense the channel idle, which is given by

$$\begin{aligned}p_1 &= \Pr\{\text{none of other } n-1 \text{ nodes access the channel} \mid \text{channel is sensed idle}\} \\ &= \dfrac{\Pr\{\text{none of other } n-1 \text{ nodes access the channel}\}}{\Pr\{\text{channel is sensed idle}\}}.\end{aligned} \tag{15}$$

If no one accesses the channel, it means that all the other $n-1$ nodes are either empty, or in sensing states but not accessing the channel. Thus, we have

$$\begin{aligned}&\Pr\{\text{none of other } n-1 \text{ nodes access the channel}\} \\ &= \left[\Pr\{\text{node is empty}\}+\Pr\{\text{node is in sensing but not scheduled to be sent}\}\right]^{n-1} \\ &= \left[(1-\rho)+\rho\sum_{i=0}^{K}\tilde{s}_i(1-q^i)\right]^{n-1} \stackrel{\text{for large } n}{=} \exp\left\{-n\rho\left[1-\sum_{i=0}^{K}\tilde{s}_i(1-q^i)\right]\right\}.\end{aligned} \tag{16}$$

The probability that the node senses an idle channel is given by

$$\Pr\{\text{channel is sensed idle}\} = \left[\Pr\{\text{node is empty}\} + \Pr\{\text{node is in sensing state}\}\right]^{n-1}$$
$$= \left[(1-\rho) + \rho \sum_{i=0}^{K} \tilde{s}_i\right]^{n-1} \stackrel{\text{for large } n}{=} \exp\left\{-n\rho\left[1 - \sum_{i=0}^{K} \tilde{s}_i\right]\right\}. \quad (17)$$

Substituting (6), (9), (12), and (13) into (16) and (17), the probability of successful transmission $p_1$ defined by (15) can be expressed as

$$p_1 = \frac{\exp\left\{-n\rho\left[1 - \sum_{i=0}^{K} \tilde{s}_i (1-q^i)\right]\right\}}{\exp\left\{-n\rho\left[1 - \sum_{i=0}^{K} \tilde{s}_i\right]\right\}}$$
$$= \exp\left\{-n\rho \sum_{i=0}^{K} \tilde{s}_i q^i\right\}$$
$$= \exp\left\{-\frac{an\rho}{D}\left[\sum_{i=0}^{K-1}(1-p)^i + (1-p)^K/p\right]\right\} \quad (18)$$
$$= \exp\left\{\frac{-a\hat{\lambda}}{p}\right\}.$$

In equilibrium, the network throughput $\hat{\lambda}_{out}$ should be equal to the aggregate input rate $\hat{\lambda}$. It is easy to show that the network throughput (14) in equilibrium can be obtained from (18). □

Note that the throughput given by (14) also agrees with that obtained by Kleinrock and Tobagi in [1] and our channel model in section 2. The consistency between these approaches indicates that it is appropriate to adopt the independence assumption among input buffers. This is because the correlation among $n$ input queues becomes weak when $n$ is large [20].

Furthermore, let random variables $S_i^*$, $F_i^*$, $F_i'^*$, and $W_i^*$ be the service completion time of an HOL packet, starting from the states $S_i$, $F_i$, $F_i'$, and $W_i$ respectively, until it is successfully transmitted. We assume, without loss of generality, that $M = 1/a$ is an integer. It is straightforward to show from the Markov chain of Fig. 4 that the generating functions $S_i(z)$, $F_i(z)$, $F_i'(z)$, and $W_i(z)$ of these service completion times can be found by solving the following set of equations:

$$\begin{cases} S_i(z) = E\left[z^{S_i^*}\right] = \alpha(1-q^i)zS_i(z) + \alpha q^i z F_i(z) + (1-\alpha)zW_i(z), \text{ for } i = 0,1,...,K \\ W_i(z) = E\left[z^{W_i^*}\right] = q^i z^M F_i'(z) + (1-q^i)z^M S_i(z), \text{ for } i = 0,1,...,K \\ F_i(z) = \begin{cases} E\left[z^{F_i^*}\right] = p_1 z^M + (1-p_1)z^M S_{i+1}(z), \text{ for } i = 0,1,...,K-1 \\ F_{K-1}(z), \text{ for } i = K \end{cases} \\ F_i'(z) = \begin{cases} E\left[z^{F_i'^*}\right] = p_2 z^M + (1-p_2)z^M S_{i+1}(z), \text{ for } i = 0,1,...,K-1 \\ F_{K-1}'(z), \text{ for } i = K \end{cases} \end{cases} \quad (19)$$

Since the service of each HOL packet always starts from the state $S_0$, the first and second moments of service time can be derived from the set of generating functions in (19) and are given in Appendix I. Note that the mean service time derived in Appendix I is consistent with the expression $\left(\tilde{f}_0 + \tilde{f}_0{}'\right)^{-1}$ given in (13). Based on the queuing model of the input buffer, we will investigate various stability issues of 1P-CSMA concerning throughput and delay in the following sections.

## 4 Stable Regions

For a 1-persistent CSMA network with $n$ nodes, suppose that there are total $n_b = \sum_{i=1}^{K} n_i$ backlogged HOL packets in the mini-slot before transmission, in which $n_i$ packets are in the sensing state of phase $i$, for $i = 1,\ldots, K$. The following HOL packets may attempt to transmit during the mini-slot of a sensing state:

- An empty node may send a newly arrived packet with probability $a\lambda$;
- An HOL packet in phase 0 will be transmitted immediately;
- A backlogged HOL packet in phase $i$ will be transmitted with probability $q^i$, for $i = 1,\ldots, K$.

Hence, the attempt rate in this mini-slot is given by

$$aG = a\lambda E[n - n_b] + \sum_{i=0}^{K} q^i E[n_i]. \tag{20}$$

The mean number of empty nodes in the system is

$$E[n - n_b] = n(1 - \rho); \tag{21}$$

while the mean number of backlogged nodes in phase $i$ is

$$E[n_i] = n \frac{\tilde{s}_i}{\sum_{j=0}^{K} \tilde{s}_j}. \tag{22}$$

It follows from (12) that the attempt rate in a mini-slot defined (20) can be given as follows:

$$aG = a\hat{\lambda}(1-\rho) + \frac{n\rho \sum_{i=0}^{K} \tilde{s}_i q^i}{\sum_{j=0}^{K} \tilde{s}_j}$$
$$= a\hat{\lambda}(1-\rho) + n\rho \frac{p+q-1}{pq + (p+q-1-pq)\left(\frac{1-p}{q}\right)^K}. \tag{23}$$

By solving equation (23), the retransmission factor $q$ can be expressed as a function of the attempt rate $G$, denoted as $q = h(G)$. For any finite cut-off phase $K$, this function $h(G)$ is on the order $1/n$, i.e., $q = O(1/n)$, which implies that the network is intrinsically unstable when the number of nodes $n$ is large. This point can be explained by considering the saturated case of Geometric Retransmission ($K = 1$) when all nodes have backlogged HOL packets. In this worst-case scenario, the probabilities of successful transmission for the Type I and Type II are $p_1 = e^{-anq}$ and $p_2 = e^{-(1+a)nq}$.

Consequently, either the retransmission factor $q$ or the probabilities of successful transmission $p_1$ and $p_2$ approach zero when the number of nodes $n \to \infty$. This inherently instable property of Geometric Retransmission has also been previously reported in [4] and [5].

On the other hand, Exponential Backoff mitigates the contention problem by pushing packets to deeper phases; therefore, the retransmission factor $q$ with Exponential Backoff ($K = \infty$) expressed in terms of $h(G)$ is on the order of $O(1)$, which suggests that the network can be stabilized even for an infinite population. A detailed study of Exponential Backoff is presented in section 5.

The characteristic equation (14) of the throughput versus attempt rate $G$ is a curve that first increases and then decreases with $G$ as plotted in Fig. 5, in which $\hat{\lambda}_{max}$ denotes the maximum throughput of equation (14). This indicates that to have optimal throughput, the attempt rate $G$ cannot be too small or too large. For a throughput smaller than the maximum throughput, $\hat{\lambda} < \hat{\lambda}_{max}$, the throughput equation (14) has two roots; the smaller and larger roots are denoted as $G_S(\hat{\lambda})$ and $G_L(\hat{\lambda})$, respectively. Considering the tradeoff between $G$ and $p$, we know $G$ should be bounded in the range between $G_S(\hat{\lambda})$ and $G_L(\hat{\lambda})$ to ensure a stable network throughput where $\hat{\lambda}_{out} = \hat{\lambda}$. Examples in Fig. 5 illustrate this point. For a given input rate $\hat{\lambda} = 0.3$ and $a = 0.1$, the network has a stable throughput $\hat{\lambda}_{out} = \hat{\lambda} = 0.3$ when $G$ is within the range [$\approx 0.347, \approx 1.981$]. Thus, a necessary condition of stable throughput of the entire system can be stated as follows:

**Stable Throughput Condition** (**STC**): *For any input rate $\hat{\lambda} < \hat{\lambda}_{max}$, the attempt rate $G$ should satisfy*

$$G_S(\hat{\lambda}) \leq G \leq G_L(\hat{\lambda}). \tag{24}$$

In general, the attempt rate $G$ is an implicit function of the retransmission factor $q$ associated with the underlying scheduling algorithm. The retransmission factor $q = h(G)$ can be obtained by solving equation (23). This functional relationship, together with the inequality (24), determines a stable region of $q$. It is easy to show that function $h(G)$ is monotonic increasing with respect to $G$. Thus, the stable throughput region $R_T$ of $q$ corresponding to the stable throughput condition (24) can be defined as follows,

$$q \in R_T = [h(G_S), h(G_L)]. \tag{25}$$

Furthermore, the network throughput is defined by $\hat{\lambda}_{out} = \min\{n(\tilde{f}_0 + \tilde{f}_0'), \hat{\lambda}\}$ and the stable throughput condition **STC** ensures that $\hat{\lambda}_{out} = \hat{\lambda}$. It follows that the **STC** implies $n(\tilde{f}_0 + \tilde{f}_0') \geq \hat{\lambda} = n\lambda$, which means the offered load $\rho \leq 1$. On the other hand, it can be shown from (13) that the offered load $\rho$ is a monotonic increasing function of the retransmission factor $q$ if the attempt rate is bounded in the range $G_S \leq G \leq G_L$, or equivalently $q \in R_T = [h(G_S), h(G_L)]$. In particular, the attempt rate $G$ will reach $G_L$ when the offered load $\rho = 1$. Therefore, if the retransmission factor $q$ is chosen from $q \in [h(G_S), h(G_L))$, the offered load $\rho$ of Geo/G/1 queue of each input buffer is strictly less than 1, which is simply the stable condition of any queuing system, that the arrival rate should be strictly less than the service rate.

As we mentioned before, the stable throughput condition (24) is not sufficient to guarantee a bounded mean delay of packets queued in each input buffer. Let $X$ be the service time of HOL packet, the mean delay is determined by the first and second moments, $E[X]$ and $E[X^2]$, of the service time. To complete the analysis of stable regions, we deduce the following additional constraint from our queuing model of the input buffer.

**Bounded Delay Condition (BDC):** *The Pollaczek-Khinchin formula for mean queueing delay $E[T]$ of Geo/G/1 queue* [21]

$$E[T] = E[X] + \frac{\lambda E[X^2] - \lambda E[X]}{2(1 - \lambda E[X])} \qquad (26)$$

*requires bounded second moment of service time $0 < E[X^2] < \infty$.*

The condition **BDC** is more restrictive than the condition **STC**. A region of the factor $q$, denoted $R_D$, that guarantees bounded mean delay can be determined by the second moment of service time $E[X^2]$. In general, the bounded delay region is a subset of the stable throughput region, $R_D \subseteq R_T$. Detailed discussions on these stable regions with the exponential backoff scheduling algorithm are provided below.

## 5 Analysis of Exponential Backoff

The exponential backoff scheme has been studied in many previous papers [4]-[8]. In this section, we discuss the stability and delay performance of this scheduling algorithm based on the conditions specified in section 4. The analytic results are verified by the simulations written in MATLAB.

### 5.1 Stable Throughput Condition of Exponential Backoff

The attempt rate $G$ of exponential backoff can also be derived from (23) and expressed as follows:

$$aG = a\hat{\lambda} + \lambda \left( q\frac{1+a-\alpha}{p+q-1} + \frac{1}{p} \right) \left[ n\frac{p+q-1}{pq} - a\hat{\lambda} \right]. \qquad (27)$$

or, equivalently, the retransmission factor $q$ can be formulated as a function of attempt rate $G$ as

$$q = h^{Ex}(G) = (1-p) \frac{A(p) + 2\hat{\lambda} + \sqrt{A^2(p) + 4a\hat{\lambda}^3(1+a-\alpha)p^2 n^{-1}}}{2\left[A(p) + \hat{\lambda} - a\hat{\lambda}^2(1+a-\alpha)p^2 n^{-1}\right]}, \quad (28)$$

where $A(p) = \hat{\lambda}(1+a-\alpha)p - a(G-\hat{\lambda})p^2 - ap\hat{\lambda}^2 n^{-1}$.

For retransmission factor $q$ in the range $0 < q < 1$, this function $q = h^{Ex}(G)$ is monotonically increasing with respect to the attempt rate $G$. It follows that the following stable throughput region $R_T^{Ex}$ of exponential backoff can be obtained from (25) and (28):

$$R_T^{Ex} = [h^{Ex}(G_S), h^{Ex}(G_L)]. \quad (29)$$

The area under the lower and upper bounds shown in Fig. 6.a is the stable throughput region $R_T^{Ex}$ of exponential backoff with $n = 10$ and $a = 0.1$, while that in Fig. 6.b is the stable region with $n = 50$ and $a = 0.1$. The maximum throughput with exponential backoff can be achieved such that $\hat{\lambda}_{max}^{Ex} = \hat{\lambda}_{max}$ as the stable region (29) shrinks to a single point when $G_S(\hat{\lambda}_{max}) = G_L(\hat{\lambda}_{max})$.

If we ignore higher order terms of $\hat{\lambda}$ and $p$ in (28), the retransmission factor $q$ is approximately equal to $1 - p$:

$$q = h^{Ex}(G) \approx 1 - p, \quad (30)$$

which implies that the stable region of exponential backoff is almost independent of the population size when $n$ is large enough. For example, the stable throughput regions for $n = 10$ and $50$ shown in Fig. 6 are very close to each other. Therefore, the stable throughput region $R_T^{Ex}$ can be approximately given as follow,

$$R_T^{Ex} \cong [1 - p(G_S), 1 - p(G_L)]. \quad (31)$$

The equality in (31) holds when the retransmission factor $q$ in (28) is equal to $1 - p$ as $n \to \infty$, which coincides with Song's results proved in [4], [5], and [8] that network throughput can be non-zero even when the number of nodes goes to infinity.

Our stability analysis is confirmed by the simulation results shown in Fig. 7 with $n = 10$ and $a = 0.1$, which closely follows the protocol details for 1-persistent CSMA. The 95% confidence intervals are shown for all the simulations points. For fixed aggregate input rate $\hat{\lambda} = 0.3$, Fig. 7 displays that stable throughput can be achieved if the retransmission factor $q$ is properly chosen from the stable throughput region $R_T^{Ex} = [0.135, 0.849]$. The simulation results match exactly with the analytical one when the retransmission factor $q$ is within the stable region. While outside the stable region, the throughput immediately drops and the size of the confidence intervals increases due to the unstable queuing behavior of input buffers.

## 5.2 Bounded Delay Condition of Exponential Backoff

The second moment of service time of the exponential backoff scheme is derived from (37) of Appendix I as follows:

$$E[X^2] = B(p_1, p_2, q) \lim_{K \to \infty} \sum_{j=1}^{K-1} \left( \frac{1-p}{q^2} \right)^j + C(p_1, p_2, q), \qquad (32)$$

where $B(p_1, p_2, q)$ and $C(p_1, p_2, q)$ are two polynomials given in (38) and (39) of Appendix I. It follows that the bounded delay condition requires the convergence of the infinite geometric sum $\lim_{K \to \infty} \sum_{j=1}^{K-1} \left( \frac{1-p}{q^2} \right)^j$, which implies

$$q > \sqrt{1-p}. \qquad (33)$$

For binary exponential backoff, i.e., $q = 0.5$, the above condition (33) is consistent with Yang's result given in [22].

To achieve stable throughput, the attempt rate $G$ should be larger than or equal to the small root $G_S$ as described in (24). Combine the condition (33) with the stable throughput region $R_T^{Ex}$ (29), and the bounded delay region is given as follows:

$$R_D^{Ex} = \left[ \sqrt{1-p(G_S)}, h^{Ex}(G_L) \right). \qquad (34)$$

The bounded delay region $R_D^{Ex}$ of exponential backoff is a subset of the stable throughput region $R_T^{Ex}$. As shown in Fig. 8, the shaded region is a bounded delay region; outside this region, i.e., $R_T^{Ex} \setminus R_D^{Ex}$, the system has stable throughput but cannot guarantee bounded delay. In this undesired region, predominating backlogged packets are pushed to deep phases with very low retransmission probabilities. If a node tries to send its HOL packet, the successful probability will be very high. Once the backlogged HOL packet is cleared, then the channel may be "captured" by subsequent packets in the input buffer, which are all in phase 0, until the queue is cleared. During this period, it appears that the network throughput is still stable, but the variance of the service time of each individual packet can be infinitely large due to this unfairness of services caused by the capture effect described in [18] and [19].

Fig. 9 and Fig. 10 illustrate the packet-queuing delay, in terms of number of packets in the system, versus the retransmission factor $q$ and the input rate $\hat{\lambda}$, respectively. Our analysis is confirmed by simulation as shown in both figures. In Fig. 9, if the retransmission factor $q$ is chosen within the region of bounded delay, then there are nearly zero backlogged packets in the system. However, if the retransmission factor $q$ lies outside the bounded delay region, the number of backlogged packets in the system would become larger even though it is still within the stable throughput region. For any retransmission factor $q$ outside the stable throughput region, the number of backlogged packets in the system immediately becomes much larger, which implies an unacceptable mean queuing delay. Other examples of the bounded delay region are shown in Fig. 10, which exhibits the number of packets in the system versus input rate $\hat{\lambda}$ for fixed retransmission factor $q = 0.2$ and $q = 0.5$. It can be clearly seen that the number of packets in the system is nearly zero within the bounded delay region. This result coincides with [5] in which the authors claimed that the exponential backoff scheme can be stable if the input rate is sufficiently small.

Since the bounded delay region $R_D^{Ex}$ specified by (30) is a subset of the stable throughput region $R_T^{Ex}$, the maximum throughput $\hat{\lambda}_{max}$ may not be achievable if $q \in R_D^{Ex}$. As shown in Fig .8, the network throughput increases with the lower bound of the bounded delay region $\sqrt{1-p(G_S)}$, and decreases with the upper bound $1-p(G_L)$ in equation (31). Therefore, the maximum throughput within the bounded delay region $R_D^{Ex}$ can be achieved when the lower bound and upper bound of the retransmission factor $q$ given by (30) are equal. That is,

$$\sqrt{1-p(G_S)} = 1-p(G_L). \tag{35}$$

In (35), we plot the respective maximum throughput within the throughput stable region and bounded delay region. For $0 < a < 1$, the maximum throughput $\hat{\lambda}_{max}$ given in Fig. 11 is always larger than that obtained from the bounded delay region. In other words, the absolute maximum throughput $\hat{\lambda}_{max}$ in the stable throughput region cannot be achieved with bounded mean delay guarantee.

# 6  Conclusion

We have analyzed the stability conditions in terms of stable throughput and bounded delay for slotted 1-persistent CSMA/CA networks. The queuing model of input buffer with the *K*-Exponential Backoff collision resolution algorithm is proposed to conduct the throughput analysis of the entire system and the performance of each individual input buffer. Based on this model, the exponential backoff scheduling algorithm has been used to establish the stable throughput region and bounded delay region with respect to aggregate input rate $\hat{\lambda}$ and retransmission factor $q$. It is shown that the network throughput of the exponential backoff scheme can always be stabilized with the proper selection of $q$; while the region shrinks remarkably subject to the bounded delay requirement. Consequently, the maximum achievable throughput of the network within the bounded delay region is slightly smaller than the absolute maximum throughput. The proposed methodology can also be applied to other CSMA-based networks, such as IEEE 802.11, in the future.

# Appendix 1. Service Time Distribution for 1P-CSMA Protocol

For exponential backoff scheme as $K \to \infty$, the first and second moments of scaled service time, $E[X]$ and $E[X^2]$, can be derived from (19) and given as follows.

The first moment:

$$E[X] = \frac{aM}{p} + \frac{aq(M - \alpha M + 1)}{q + p - 1}. \tag{36}$$

The second moment:

$$E[X^2] = B(p_1, p_2, q) \lim_{K \to \infty} \sum_{j=1}^{K-1} \left(\frac{1-p}{q^2}\right)^j + C(p_1, p_2, q), \tag{37}$$

where

$$B(p_1, p_2, q) = 2a^2 (M + 1 - \alpha M)^2 \left[ \frac{(1-p)}{p^2 q^2} - \frac{q^2}{(p + q^2 - 1)} - \frac{(1-p)^2 (1+q)}{qp(p + q^2 - 1)} \right], \tag{38}$$

and

$$C(p_1, p_2, q) = a^2 \left\{ \frac{(M-1)M}{p} + \frac{q(M - 2 + 2\alpha M)(M + 1 - \alpha M) - \alpha q M}{(p + q - 1)} + \frac{2q^3 (M + 1 - \alpha M)^2}{(p + q^2 - 1)(p + q - 1)} \right. \\ \left. + \frac{2(1-p)(M + 1 - \alpha M)qM}{p(p + q - 1)} + 2\left[ M - \alpha M p - (1 - \alpha) p_2 M \right] \left[ \frac{M}{p^2} + \frac{q(M + 1 - \alpha M)}{(p + q - 1)^2} \right] \right\} + E[X]. \tag{39}$$

# List of Figure

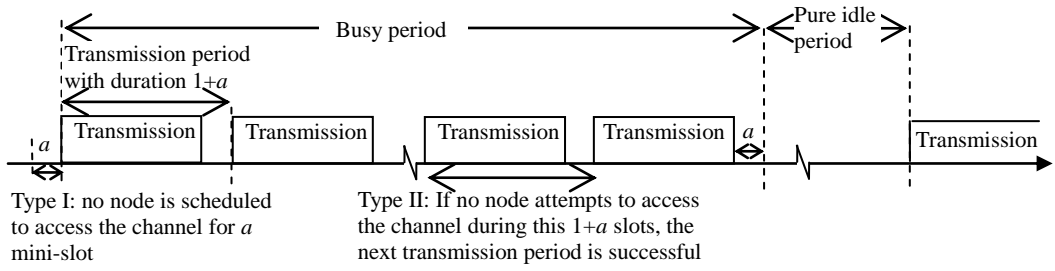

Fig. 1.  The renewal process of 1P-CSMA with two types of transmission periods.

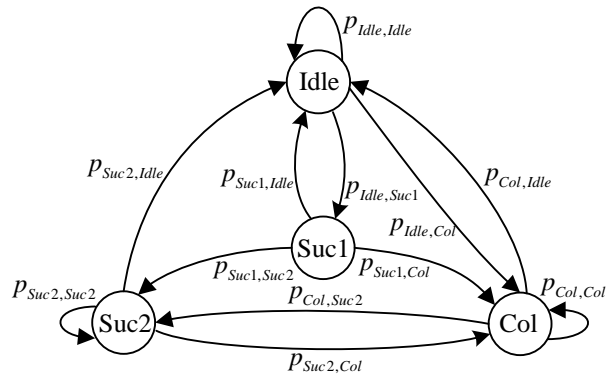

Fig. 2.  Markov chain of the 1P-CSMA channel.

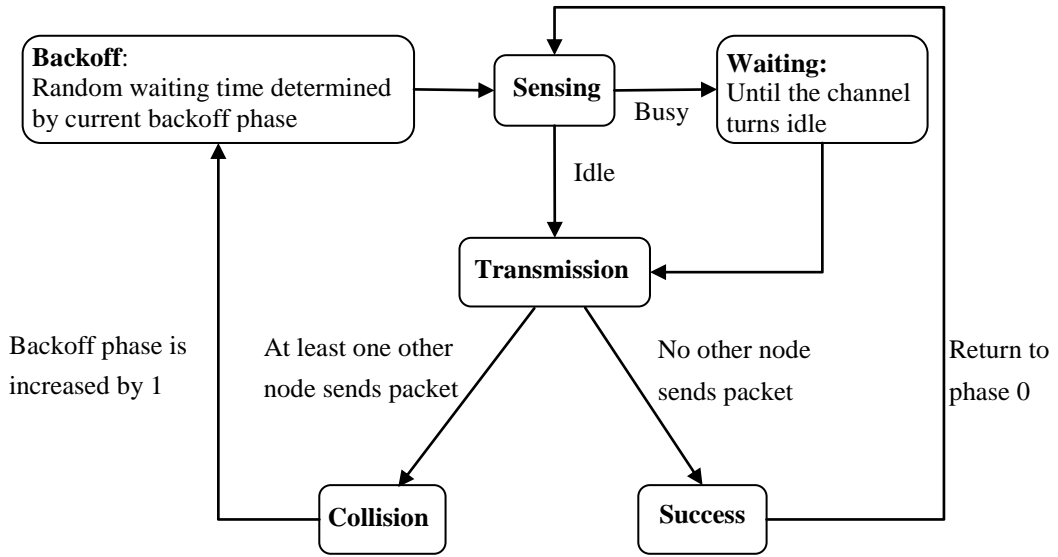

Fig. 3.  Flow chart of access behaviors for HOL packets.

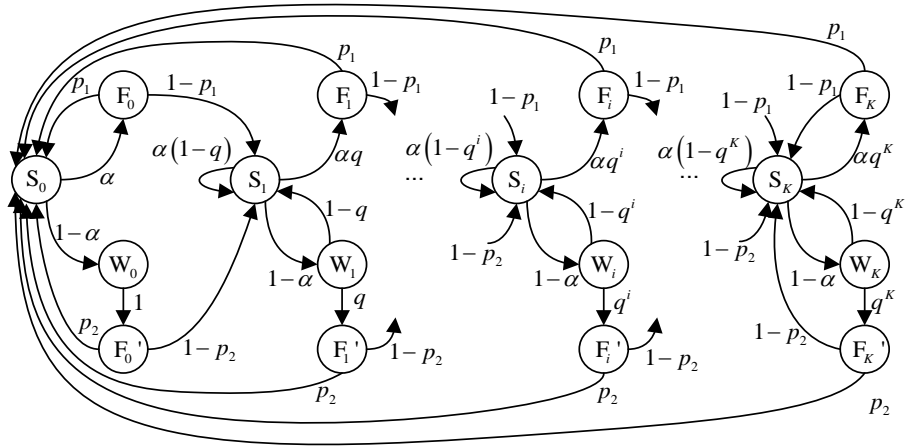

Fig. 4. Markov chain of HOL packet.

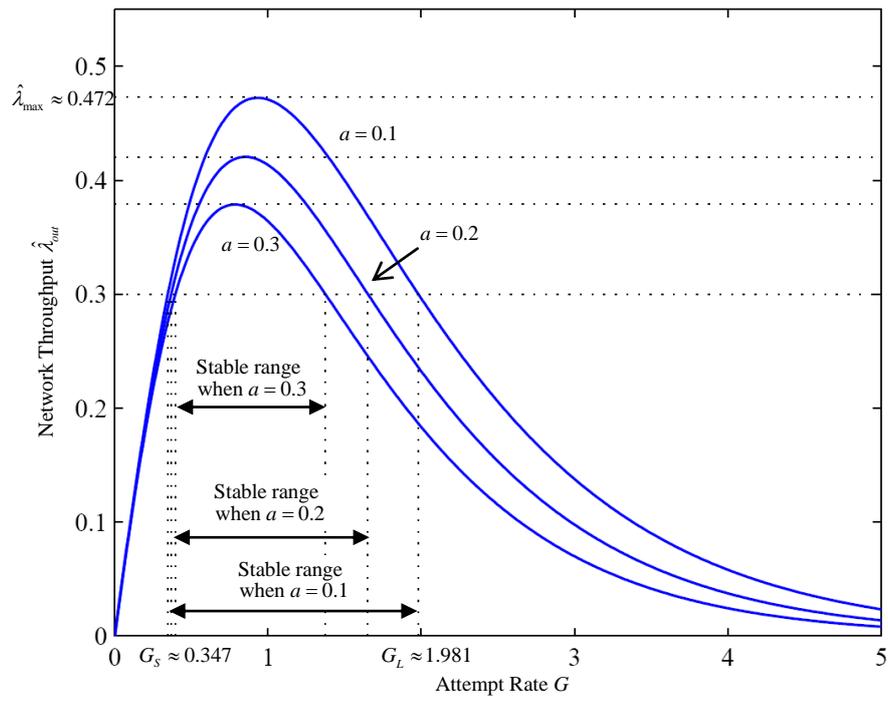

Fig. 5. Throughput versus attempt rate for 1P-CSMA.

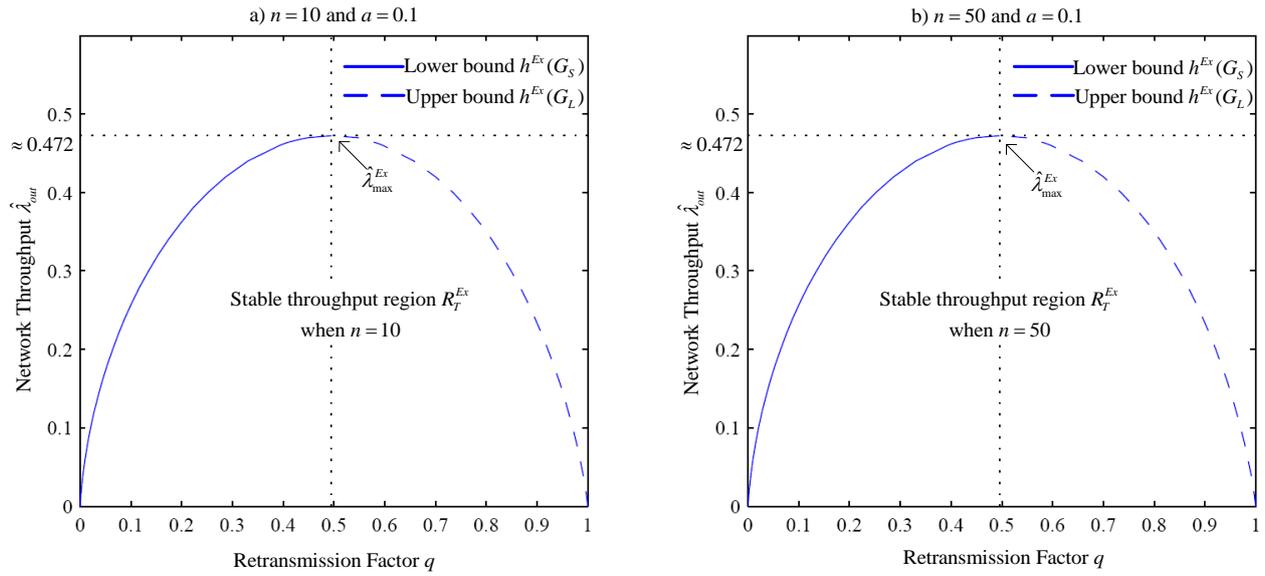

Fig. 6. Stable throughput regions of exponential backoff: a) $n = 10$ and b) $n = 50$.

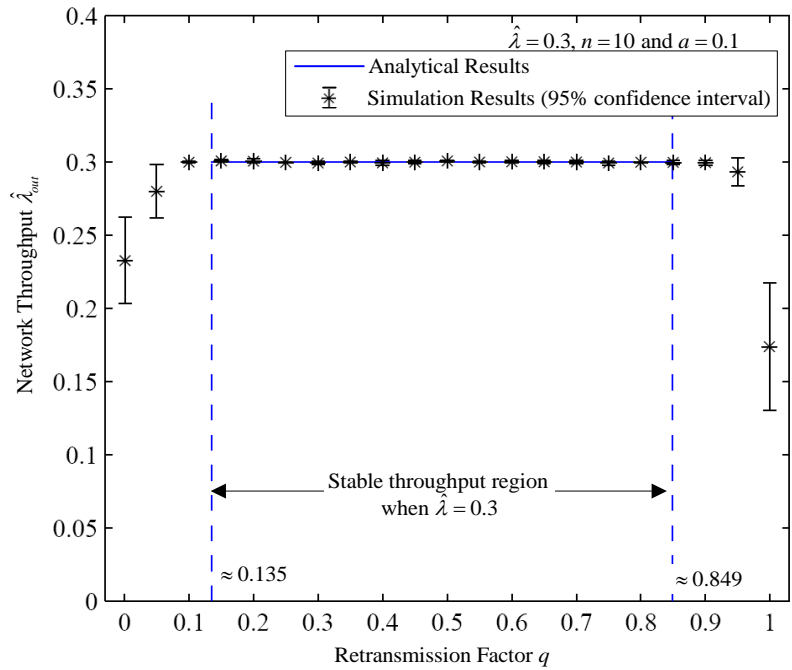

Fig. 7. Simulation results of throughput for exponential backoff with fixed input rate.

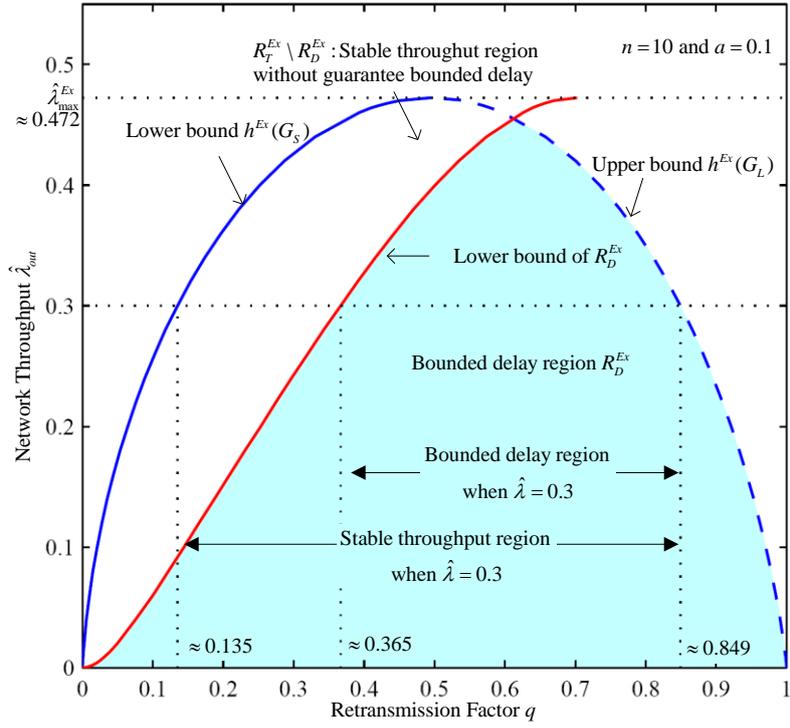

Fig. 8. Stable throughput region and bounded delay region of exponential backoff.

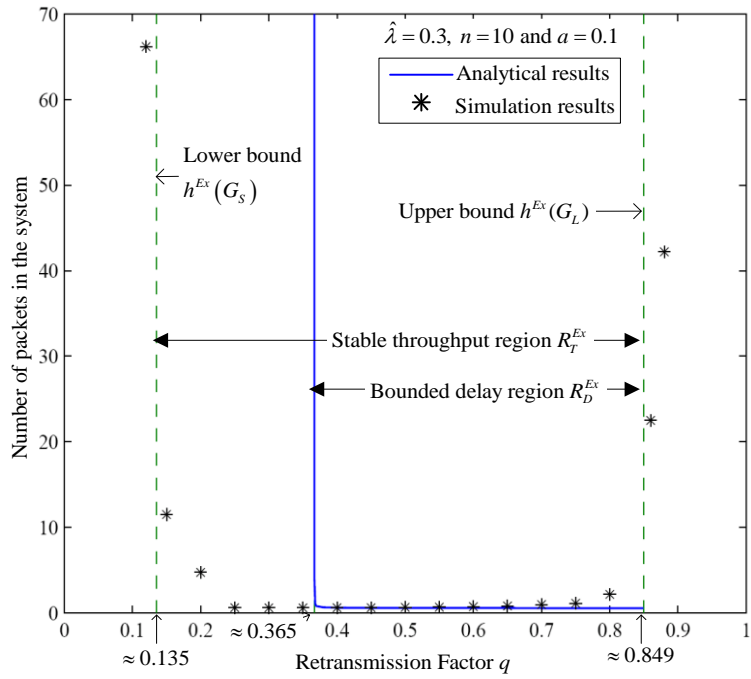

Fig. 9. Number of packets in the system versus retransmission factor $q$ for exponential backoff.

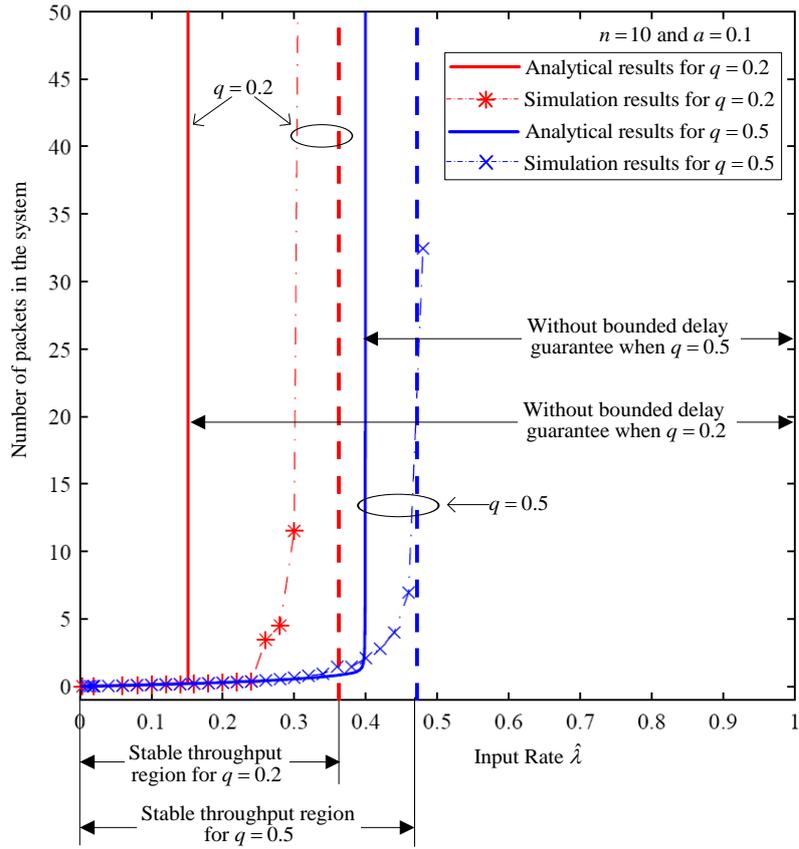

Fig. 10. Number of packets in the system versus input rate for exponential backoff.

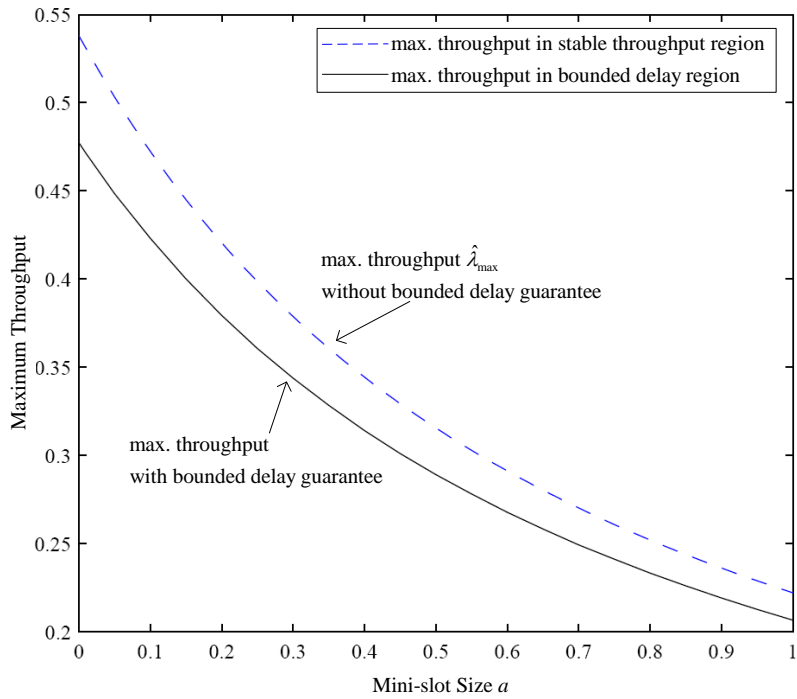

Fig. 11. Maximum throughputs in stable throughput and bounded delay regions.